\documentstyle[epsf]{aipproc}

\begin{document}
\title{
A  Dynamical Thermostat Approach To Financial Asset Price Dynamics 
} 

\author{Stefan Thurner}
\address{Institut f\"ur Mathematik NuHAG and Klinik f\"ur HNO, 
Universit\"at Wien, \\A-1090 Vienna, Austria 
}
\maketitle
\begin{abstract}
A dynamical price formation model for financial assets is presented. 
It aims to  capture the essence of speculative trading 
where mispricings of assets are used to make profits. 
It is shown that together with the incorporation of the 
concept of risk aversion of agents the model is able to reproduce 
several key characteristics of financial price series. 
The approach is contrasted to the conventional view of 
price formation in financial economics.  
\end{abstract}

\section*{Introduction}
A remarkable feature of (financial) economics is that an 
immensely complex system - consisting  of a  relatively large 
number of human beings, interacting through the exchange of 
goods, information, services,  and money - reduces to one-dimensional 
functions in time, the prices of assets. 
These prices potentially carry substantial  information about  
the underlying system, and are therefore an intriguing subject 
to study. 

Technically   
the asset price,  be it of a stock, option, or any other liquid 
financial derivative  
can change whenever a trade (interaction of market participants) 
takes place.
Prices are formed through the actions of three types of agents: 
A seller offers an asset at a certain price, usually through a 
sell order to the market maker. Potential buyers  
offer to buy the asset at a certain price, or at any price.  
The market maker sets the actual unique price at which 
trades then  occur. 
For convenience one frequently works with the log-return 
process rather than with the price process itself. 
The log-return over a specified time interval (aggregation time $\tau$) 
is simply given by the logarithm of the quotient of prices 
separated by time $\tau$:
\begin{equation}
r_{\tau}(t) = {\rm log}( p(t+\tau) ) - {\rm log}( p(t) ) \quad  . 
\label{logreturn} 
\end{equation}
Due to the addition property of logarithms
it is easy to change from a given  aggregation time 
$\tau$ to higher ones by taking
the prices which form the boundary of the time interval of interest. 
These return series possess a highly non-trivial structure, and 
over the last decades considerable effort has been devoted  
to quantitatively characterize such  processes.  

Maybe the best known fact about  financial price time-series is that 
the distribution of returns is leptocurtic. On short time-scales 
($\tau$ ranging from minutes to several days),  for small 
price fluctuations the return distribution can be well approximated by a 
L{\'e}vy stable distribution \cite{Mandelbrot63}, 
while its tails - emerging from violent 
price changes - generally follow a power law, see e.g. \cite{Koedijk90}. 
It is well established that for longer time-scales, i.e., 
aggregation times $\tau$ of  three weeks and longer, the distribution  
functions of returns are almost Gaussian. 
Historically there has been some confusion on this subject: 
In the 1960s Mandelbrot and Fama found 
empirical evidence 
that the  cumulative distribution
functions of returns  of time-scale $\tau$, $f(r_{\tau})$,
behave like  stable  L{\'e}vy distributions, which 
are characterized by a
parameter $1\leq \mu \leq2$, where $\mu=2$ corresponds to the special case of
a normal distribution.
For $\mu < 2$ the distributions are fat tailed and higher momenta
- including the second  - are infinite.
They estimated an empirical value of $\mu=1.7$,  indicating that
higher momenta will not exist and therefore putting in doubt most
statistical quantities.
In the 1980s power laws for the tails of the distributions have been 
found, i.e. $f(r_{\tau}) \sim |r_{\tau}|^{-\alpha}$,
with $\alpha > 2$. This result has been confirmed
throughout the 1990s and the exponent is nowadays believed to be 
close to $\alpha = 3$ \cite{Plerou99}.

For liquid assets, ${\cal O}(100 -1000)$  trades per day,  
the stock returns are basically uncorrelated, 
i.e., the autocorrelation functions with time lags larger than one are 
usually compatible with zero. However, one notices the fact that 
there are periods in time when price fluctuations are stronger than 
at other times. This is called volatility clustering and can 
be captured quantitatively by the autocorrelation  function of the 
absolute values (or squares) of the returns. These volatility 
autocorrelations 
do not vanish exponentially but again show an empirical 
power-like decay, ${\rm lag}^{-\beta}$, with $\beta \sim 0.1-0.5$,   
see e.g. \cite{Ghashgaie96,Arneodo98,Ding93}.    

There have been observations that higher momenta of 
volatility $\langle |r_{\tau}|^q\rangle $ 
are scaling processes with respect to aggregation time $\tau$ 
and that the 
corresponding scaling exponents are not linear functions 
in $q$, which is an indication for a multiscaling 
process \cite{Ghashgaie96,Schmitt99}.  

%
Further empirical features of price processes are that 
prices are non-stationary, the  second moment of returns 
$\langle r^2 \rangle$ exists (but converges slowly), whereas the 
fourth  moment $\langle r^4 \rangle$ does not. 
Prices are usually quantized (1/16 dollar at the NYSE), and can not 
jump arbitrarily much from one trade to the other. 
The closing price is the price of the asset at the 
end of a trading day.

In this paper we model the one-dimensional price process, 
by casting some fundamental characteristics of economic systems 
and human behavior into a set of differential equations, and 
compare the emerging  price time-series statistics with known facts of 
real price series. The paper is organized in the following way: 
The next section is a mini-overview on
the standard views in (micro) economics of price formation. 
We then motivate and propose a dynamical system, show 
some of its features,  and discuss the results.  
 
\section*{Standard treatment in economics}
Financial economics literature, in particular the field of 
micro-economics, is dominated by several concepts which   
are powerful from a theoretical standpoint but which are 
known to have their limits when it comes to the analysis of 
real time-series. These concepts are: 

{\it Dogma 1:}
Prices are the equilibrium result of trades, i.e., all sellers
and buyers come to an optimum state (Pareto optimum), where
everybody is better off than without doing the trades. It can be shown 
quite generally with the help of the Brower fixed point theorem, 
that  such an equilibrium state exists  
which also provides an optimum \cite{Arrow54}. 
For this to be true it is important 
that the influence of trading barriers is kept small.   
In this approach all the agents involved in trading  maximize 
their von Neumann-Morgenstern utility functions. These are monotonically 
rising convex functions, which describe the ``happiness'' 
of the individual agent as a function of his wealth. 
The fact that utility functions rise monotonically reflects 
that ``more is preferred to 
less'', the convexity takes into account the risk-aversion 
of agents. 
The main problem associated with this view is that it is  not able 
to account 
for speculative trading at all. On the other hand one knows that 
speculative trading is dominating entire markets. For example 
in 1995 the daily trading volume at the foreign 
exchange markets exceeded $10^{12}$ USD which was about 20 times 
the daily world gross national product.   

{\it Dogma 2:}
Markets are efficient. The basic requirement for efficiency is that 
all participating agents are fully rational, that there is rapid 
information flow, there are no transaction costs or other market friction, 
and that everybody has identical access to all the relevant information.  
As a consequence,  prices simply reflect the expectations on 
future earnings from investments made today. Thus trading would  only 
occur at the appearance of news which would influence those 
expectations. Other than that there would be no trading activity and 
speculative trading would be entirely useless. 
Another consequence of the efficiency hypothesis is that 
there exist no patterns in financial time-series which could be 
exploited in one or the other fashion. 
Also there should not exist arbitrage (risk-less profit) 
opportunities. In fact the non-existence of arbitrage forms the basis 
of many celebrated results in financial mathematics, especially 
in the field of pricing financial derivatives. 
 
Obviously the efficient market hypothesis is rather unrealistic 
and there are strong problems related with it:  
Agents are {\it not} fully rational, but often tend to have non-rational 
strategies, such as heard behavior, believe in experts' opinions, 
etc.    
Moreover it has been clearly demonstrated that most 
trading is {\it not} influenced by news \cite{Cuttler89}. 
Finally, it is needless to mention that there exist financial arbitrage 
companies which are specialized in 
exploiting  patterns in financial time-series. 

{\it Dogma 3:}
Mathematical tools: In financial literature the most commonly 
used mathematical tool is  game theory. Most of the time, non-iterative 
one or two 
period games are considered, which severely limits any  
study of the dynamics of a system.  
If dynamics is studied, usually the corresponding models are 
based on Wiener processes (Brownian motion). 
For example as a starting point for many models a price process 
is assumed to have the form 
\begin{equation}
d p(t) = \mu \, p(t) \, dt   + \sigma \, p(t) \, dW(t)  \quad , 
\label{BSdyn}
\end{equation}
with $\mu$ the so-called drift term, $\sigma$ the volatility, 
and $W(t)$ the Wiener process. The first term on the right side 
determines the overall exponential growth of an asset price with 
growth-rate $\mu$. Superimposed on that, the second term 
introduces a source of stochasticity, whose relative importance is 
controlled by the volatility $\sigma$. 

Wiener processes are nowadays mathematically well understood 
and are relatively easy to handle, since they are Gaussian processes. 
These kind of processes have led to a number of celebrated 
results in option pricing\cite{BS}, interest rate models, etc. 
For an overview see \cite{Bjoerk99}. 
However, there are severe problems associated with Wiener processes 
if one tries to explain real price series. 
Clearly, in Gaussian price processes no power laws are  present in the 
return distributions, there is no volatility clustering  and 
there exist no power laws in the volatility autocorrelations. 

In order to make things more  realistic, so-called ARCH-GARCH processes 
have been introduced \cite{Bollersev86}. 
Such models are nowadays frequently used in everyday eco\-no\-me\-trics 
and risk management.
A GARCH(1,1) process 
has  one source of randomness and obeys the following evolution
equations:
\begin{eqnarray}
\epsilon_t & = & \sigma_tz_t \quad ; \quad z_t 
         \sim {\cal N}(0,1) \nonumber \\
\sigma_t^2 &=& a \sigma^2_{t-1}+b \epsilon^2_{t-1} +c \quad , 
\end{eqnarray}
where $z_t$ are uncorrelated normal random variables, and $a,b$ and $c$ 
are real numbers.
Such models are able to explain 
clustered volatility, but fail to give power laws in the 
volatility  autocorrelations. The main problem with GARCH models 
is that they are ad hoc models and do not relate the variables to an  
economic context. 

Obviously there is a need for a better understanding of the 
origin of the basic features of price processes.  
Recently there have been put forward some new ideas from 
the mathematics and physics community, some  of which are 
certainly more successful in describing reality 
than the standard methods used in eco\-no\-mics.  
Just to mention a few directions of new developments,  
there are contributions from
evolutionary game theory \cite{Siegmund}, 
agent based models and minority games \cite{Artur}, and  
spinglass-inspired  models \cite{Iori}.  
Stochastic models of market maker behavior have been discussed  
in  \cite{Maslov} and in market impact models  \cite{Farmer00} 
several realistic price formation scenarios are studied.   
In the following we present a dynamical systems approach, which tries to 
model fundamental behavior of speculating market participants. 
The aim is to capture this behavior in a set of coupled 
differential equations. The solution to this system yields 
the price process. 

\section*{The dynamical thermostat model}
The model is supposed to capture three fairly general features of 
price dynamics and investor behavior: 
 
{\it Fact 1:} 
If money is invested in a bank or in government bonds, which is 
often referred to as a risk-less investment, wealth   
will increase exponentially with a fixed interest rate 
(for short-term investments). If one decides 
to invest in risky assets like stock, 
this exponential tendency (drift) should also be present  
with a somewhat larger expected rate of return, which compensates 
for taking the risk (risk premium). 
The price dynamics of a risky asset will be more complicated 
since it is coupled to a complex dynamical system - the market. This can be 
written as 
\begin{eqnarray}
\label{modelA}
\dot p(t) & = & \mu  \, p(t) - c \, \xi_1(t) \, p(t)  \quad , 
\end{eqnarray}
with $\mu$ the drift coefficient and $c$ a coupling constant. 
$\xi_1$ is a coupling or friction variable, whose dynamics is 
the main subject of this work. 
Note that this equation is formally similar to the standard price 
dynamics using Wiener processes, Eq. (\ref{BSdyn}).   

\begin{table}[b]
\caption{Meaning of model constants and variables and its relation to physics} 
\begin{tabular}{lll}
symbol & meaning in model & relates to in physics \\
\hline
$p$& price function  & momentum \\ 
$\mu$& risk-free rate (bank rate) & ``force'' term   \\
$c$& primary coupling of the asset to the market & \\
$\xi_1$& relative price-change variable & thermostat variable\\
$\xi_2$& control variable for $\xi_1$ & thermostat variable\\
${\cal T}_0(t)$& fundamental or ``true'' value of asset & temperature  \\
${\cal T}_1(t)$& collective  (inverse) risk-aversion factor  & temperature \\
$\tau_i$ &  response on-set times & response on-set times
\label{Var}
\end{tabular}
\end{table}

{\it Fact 2:} 
The essence of speculative trading is that agents 
try to make profits by  using  mispricings of assets. 
Suppose there exists the ``true'' value of an asset 
at a given time $t$, $ {\cal T}_0(t)$. 
The actual price $p(t)$ can, and in general will,  
differ from this value. 
As an example of how such a mispricing can be used to make profits
imagine that the asset is ``under-priced'', i.e., the actual price is  
lower than the true value,  $ p< {\cal T}_0(t)$. 
Alert agents who notice this difference  
will buy the asset and hold it until the market as a whole 
(not that alert) values the asset correctly. 
In this course the price  of the asset will 
rise by an  increased demand,  and the 
mispricing gets reduced. 
Obviously the rise in price can be used by the agents to realize profits. 
If an asset is ``over-priced'' there are also  ways to make profits 
by so-called ``short-selling''. 
In this context it seems reasonable to view the price changes $\xi_1$ in 
Eq. (\ref{modelA}) proportional to the mispricing: 
\begin{eqnarray}
\dot \xi_1 & = & \frac{1}{\tau_1^2} 
\left[ p - {\cal T}_0(t) \right] \quad ,  
\label{modelB}
\end{eqnarray}
with $\tau_1$ being a constant, which can be interpreted 
as the on-set time of the correctional movements. 
As long as the asset is under-priced, $\dot \xi_1$ is negative, $\xi_1$ 
will decline and eventually fall below zero, which leads to a positive 
second term on the right side of Eq.  (\ref{modelA}), which makes the 
price rise. 

{\it Fact 3:} 
Price changes are not un-restricted, which can be explained by the 
fact that agents are risk averse. 
Agents are aware that they can 
misjudge the situation about mispricings  
and as a consequence make losses.  
Since they are not entirely sure if their investment will be  
profitable or not, they will invest only limited amounts.  
The model should thus have a handle to restrict the price 
changes $\xi_1$ to a region around zero with  controllable 
variance. Technically this can be done by introducing a new 
dynamical variable $\xi_2$ and by adding a term $ - \xi_1 \xi_2 $ 
to the left hand side of Eq. (\ref{modelB}).  The dynamics of 
$\xi_2$, which keeps the variance of $\xi_1$ at a value of 
${\cal T}_{1}$,  is  given by 
\begin{eqnarray}
\dot \xi_2 & = & \frac{1}{\tau_2^2} \left[ \xi_{1}^2 - 
{\cal T}_{1}(t) \right]    \quad, 
\label{modelC}
\end{eqnarray}
where $\tau_2$ is another time constant.  ${\cal T}_1(t)$
is a factor (maybe time dependent) that can be 
interpreted as a measure of collective 
inverse risk aversion of the involved agents. 
If ${\cal T}_{1}$ is small, agents are not willing to take risk
and only small price changes will occur. 
\begin{figure}
\begin{tabular}{ll}
\vspace{-5mm} 
\epsfxsize=7.0cm\epsffile{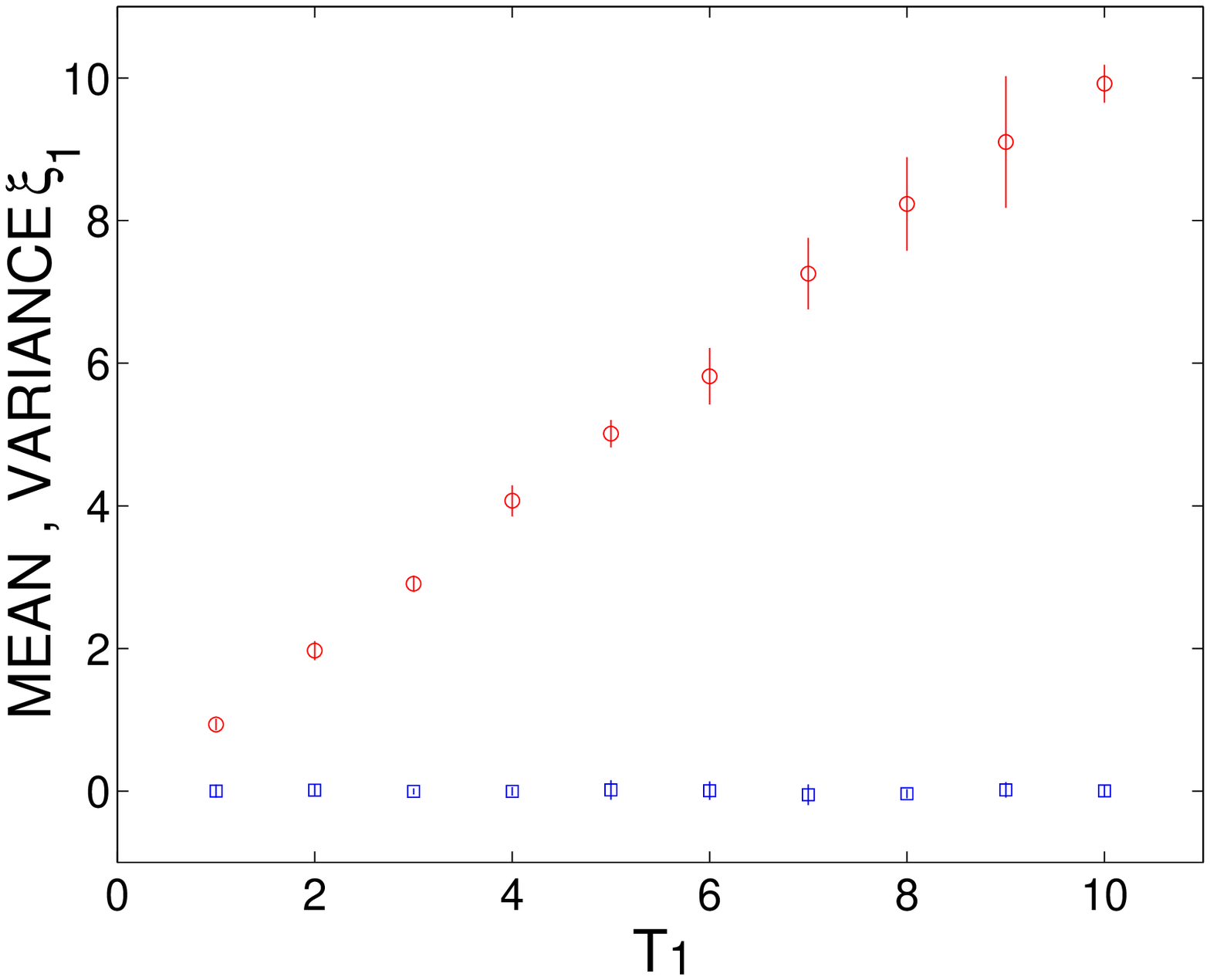} & 
\epsfxsize=7.0cm\epsffile{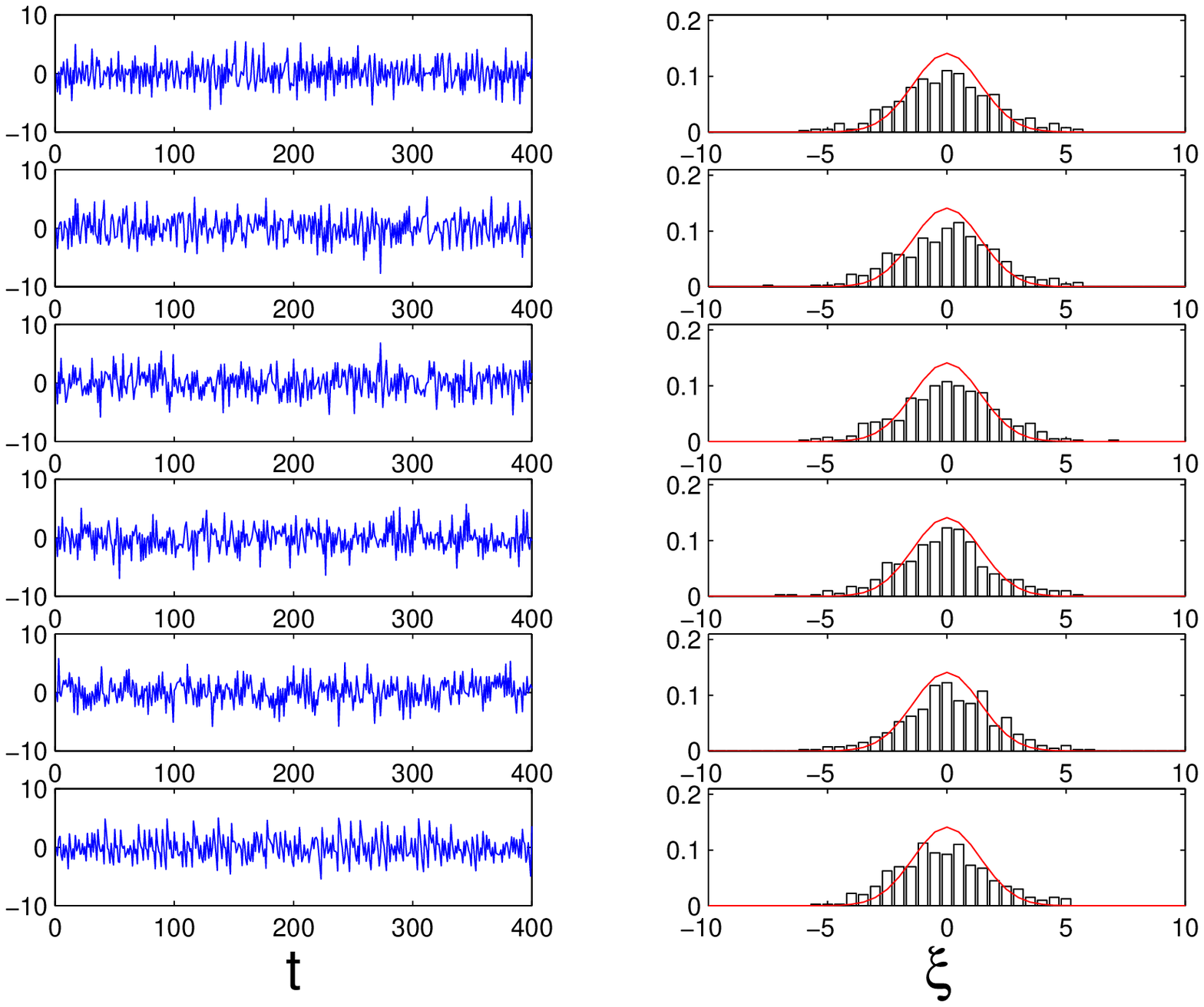}  \\
\hspace{9mm}{\large (a) } & \hspace{2mm} {\large (b) } \\
\end{tabular}
\caption{(a) Mean (boxes) and variance (circles) of $\xi_1$ as a 
function of ${\cal T}_1(t)=1,2,\cdots,10$. In (b) several 
time-series of $\xi_1(t)$ are shown with the corresponding histograms, 
which indicate fat tailed distributions. }
\label{3d}
\end{figure} 
By gathering the above arguments we propose the following model: 
\begin{eqnarray}
\dot p & = & \mu \, p -c \, p \, \xi_1 \nonumber \\
\dot \xi_1 & = & \frac{1}{\tau_1^2} \left[ p- {\cal T}_0(t) \right]  -\xi_1 \xi_2 \nonumber \\
\dot \xi_2 & = & \frac{1}{\tau_2^2} 
\left[ \xi_1^2- {\cal T}_1(t) \right]  \quad  . 
\label{model}
\end{eqnarray}
The variables are collected in Table \ref{Var}. The first 
two equations have the form of dynamical thermostat equations which have 
been introduced some time ago \cite{Nose}. 
To build the bridge to statistical physics, if one considers $p^2$ 
instead of $p$ in Eqs. (\ref{model}),  the first equation 
stays the same up to a rescaling factor 2, and the second equation 
can be seen as a dynamical  thermostat, which keeps the kinetic energy 
($p^2$) at a temperature ${\cal T}_0(t)$. $\tau_1$  
is then the thermostat on-set time. 

\subsection*{Model features}
The model is written in continuous variables. However, 
trading is not a continuous process but occurs at 
specific points in time.  
To capture this feature 
we solve the equations numerically with a Runge Kutta 4th order solver 
with time increments $\Delta t$ of 1/1000 - 1/10000. 
Every time increment  is supposed to model one single 
trade (tic-scale). 
The prices are recorded at integer times 1,2,...,N which represent the 
``closing prices'', i.e., one time unit is 1 day.  
$\Delta t$ can thus be considered an additional (hidden) 
parameter of the model. In the following we decided to use 
a dynamical step size, according to predetermined 
error tolerances.  The variable of interest is $p$, 
which will be analyzed in the following in the identical 
\begin{figure}[htb]
\begin{tabular}{ll}
\vspace{-52mm} 
\epsfxsize=7.0cm\epsffile{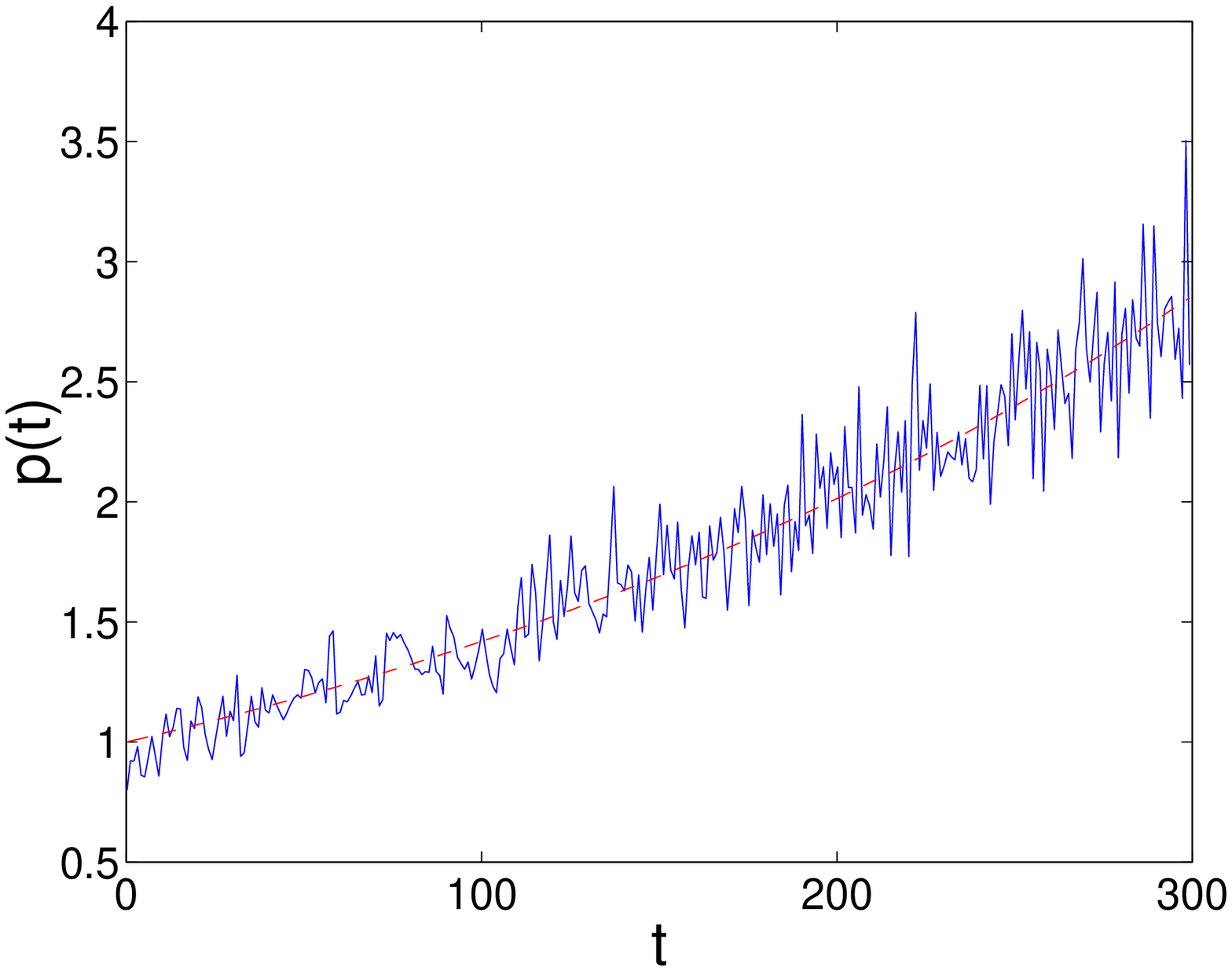} &  
\epsfxsize=7.0cm\epsffile{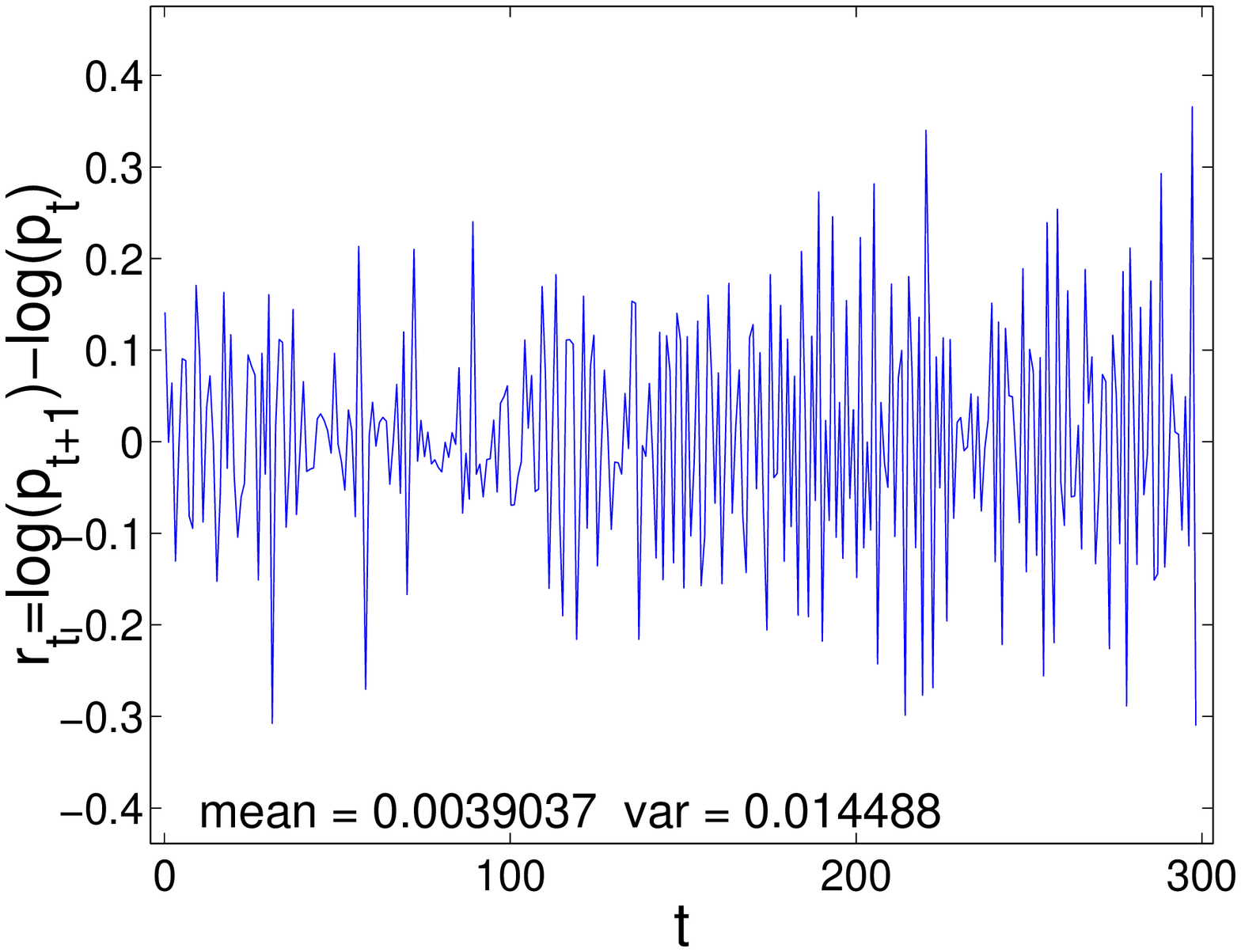} \\
\vspace{44mm} 
\hspace{12mm}{\large (a) } & \hspace{12mm} {\large (b) } \\
\vspace{-53mm} 
\epsfxsize=7.0cm\epsffile{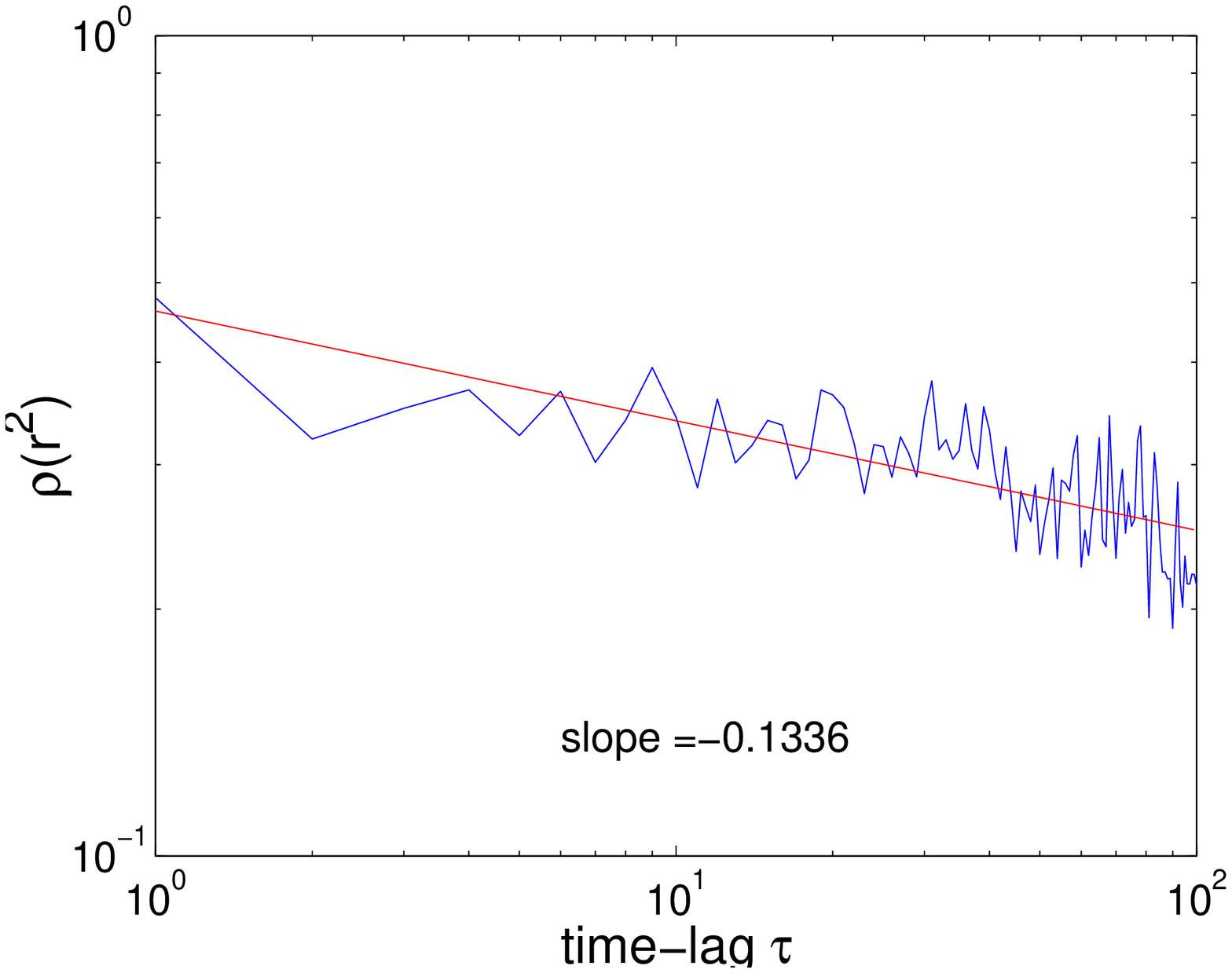} & 
\epsfxsize=7.0cm\epsffile{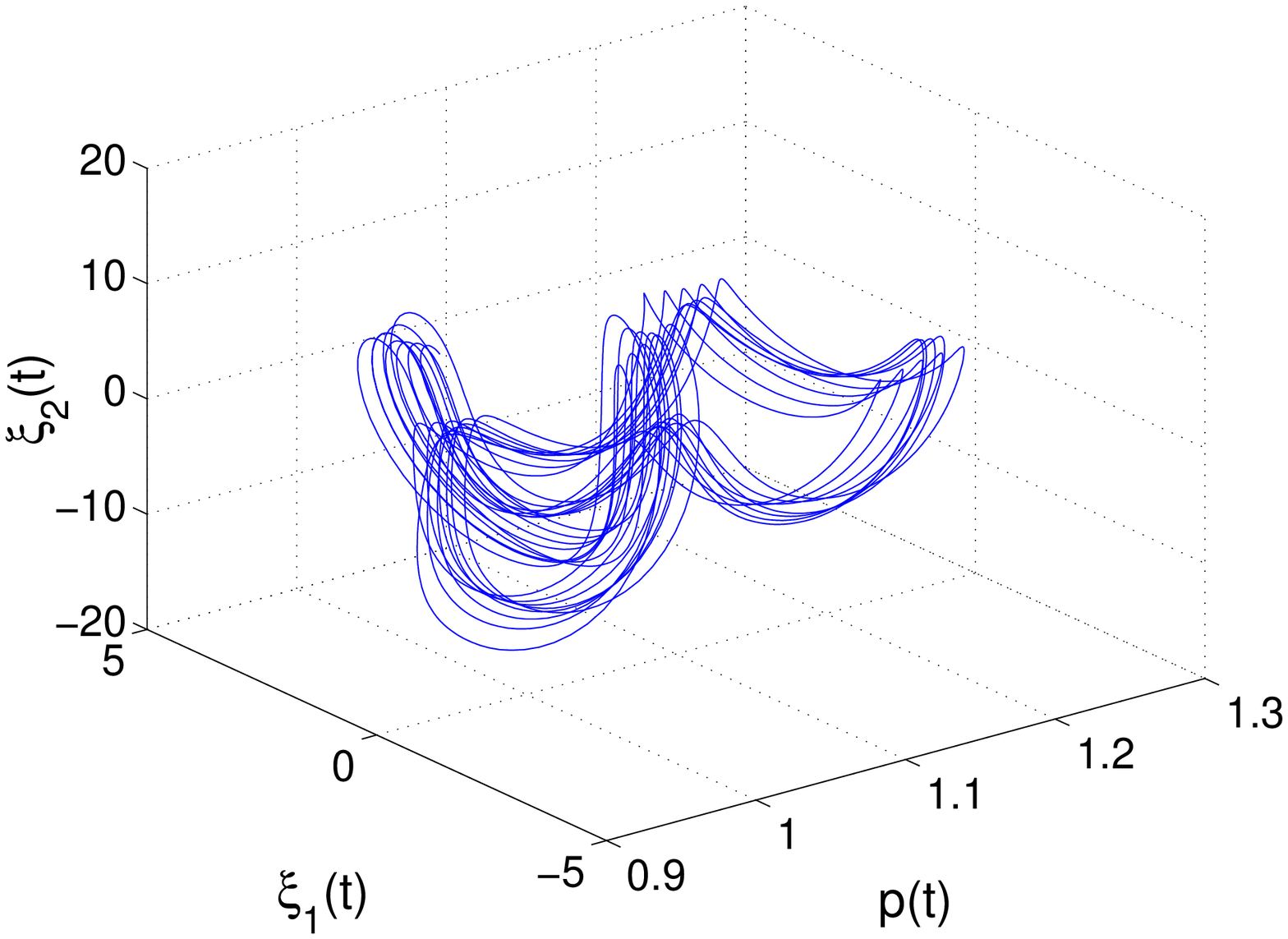}\\
\hspace{12mm}{\large (c) } & \hspace{12mm} {\large (d) } \\
\vspace{42mm} 
\end{tabular}
\caption{(a) Model price series for 300 time steps, following 
a ${\cal T}_0(t)={\rm exp}(rt)$. (b) The corresponding return process 
exhibits volatility clustering which leads to non-trivial 
volatility autocorrelation functions which are compatible with a 
power decay, (c). In (d) a part of the trajectory of the same process 
is shown. }
\label{resA}
\end{figure} 
manner as realistic stock data.     
In Fig. \ref{3d}a  we show by numerical simulation 
that the third equation in 
Eqs. (\ref{model}) controls  the variance of price changes, such that  
 ${\cal T}_1$=Var($\xi_1$).  
The mean values of $\xi_1(t)$ (boxes) are clearly compatible with zero, 
the variance (circles) rises linearly with the (constant) 
value of ${\cal T}_1$. The corresponding distributions of the $\xi_1(t)$
appear to be fat tailed but have the desired variance. 
The time-series and corresponding histograms of friction variables $\xi_1$ 
in a generic 
case are seen in Fig. \ref{3d}b. The solid line
in the histograms is ${\rm exp} [- (\xi_1/ {\cal T}_1)^2] $.

\begin{table}
\caption{Parameter dependence of log-return distribution functions.
$\alpha$ corresponds to the tail exponent at a given aggregation time $\tau$.
For all simulations we took 1000 time steps, and fixed 
${\cal T}_1={\cal T}_0$. 
} 
\begin{tabular}{ c c c c c  }
 c  & 
$\tau_1=\tau_2$  (response)  & 
$\langle r_{\tau=1}   \rangle$ &  
$\langle r_{\tau=1}^2 \rangle$ &  
$\alpha_{\tau=1}$ \\
\hline
0.18 & 0.3 & 0.0038 & 0.037 & 2.49 \\
0.18 & 0.4 & 0.0041 & 0.051 & 3.25 \\
0.18 & 0.5 & 0.0034 & 0.131 & 2.27 \\
0.18 & 0.6 & 0.0040 & 0.097 & 2.12 \\
0.30 & 0.3 & 0.0036 & 0.067 & 2.62 \\
0.30 & 0.4 & 0.0039 & 0.064 & 3.07 \\
0.30 & 0.5 & 0.0039 & 0.081 & 1.60 \\
0.30 & 0.6 & 0.0037 & 0.120 & 1.36   
\end{tabular}
\label{ret}
\end{table}
For simplicity we set $\mu=0$ for the remainder of the paper. 
To model a realistic situation, we chose ${\cal T}_0(t)={\rm exp}(rt)$ 
with $r=0.0007$ which corresponds to an average growth of about 19\% per 
year ($\sim250$ trading days). We assume that the risk aversion 
of agents becomes less as their wealth increases (convexity of 
utility functions), which we took into account by simply 
setting ${\cal T}_0(t)={\cal T}_1(t)$. Certainly one could think of 
other choices which capture the same feature.  
The on-set times $\tau_1$ and $\tau_2$ 
have been chosen to be less than one, which states that decisions on 
mispricings are made on a time-scale below one day, which is nowadays 
a realistic assumption for liquid assets (intra day trading).  
With the coupling constant $c$, it is 
possible to weigh the fluctuations relatively to the drift term. In the 
following we set it around 0.1 and 0.3.   
In Fig. \ref{resA} we show model results for the price series (a) 
and the corresponding log-returns (b).  Clearly,  the price follows the 
exponential form of ${\cal T}_0(t)$ (dashed line) 
and the returns show clusters of enhanced volatility.  
 
To become more quantitative we looked at the first 100 lags of the 
normalized autocorrelation functions of squared log-returns.  In 
Fig. \ref{resA}c the situation is shown for the same parameters as before. 
The first 100 lags of the autocorrelation is seen to be compatible 
with a power decay (straight line) with an exponent 
of the order of $\beta \sim 0.1$, which is within the realistic range. 
 
In Table \ref{ret} we gather the power exponents 
$\alpha$ of the return distribution functions for an aggregation 
time of $\tau=1$ ``day''.    
These values have been obtained by linearly fitting the tails of the 
distribution functions in a log-log plot, and by averaging over 3 
independent runs.  In most cases $\alpha$ lies in the range 
between 2 and 3, depending somewhat on the on-set times $\tau_1$ and 
$\tau_2$. This result is again in good agreement with observed data. 

\begin{figure}
\begin{tabular}{cc}
IBM Stock & Model \\
\epsfxsize=7.1cm\epsffile{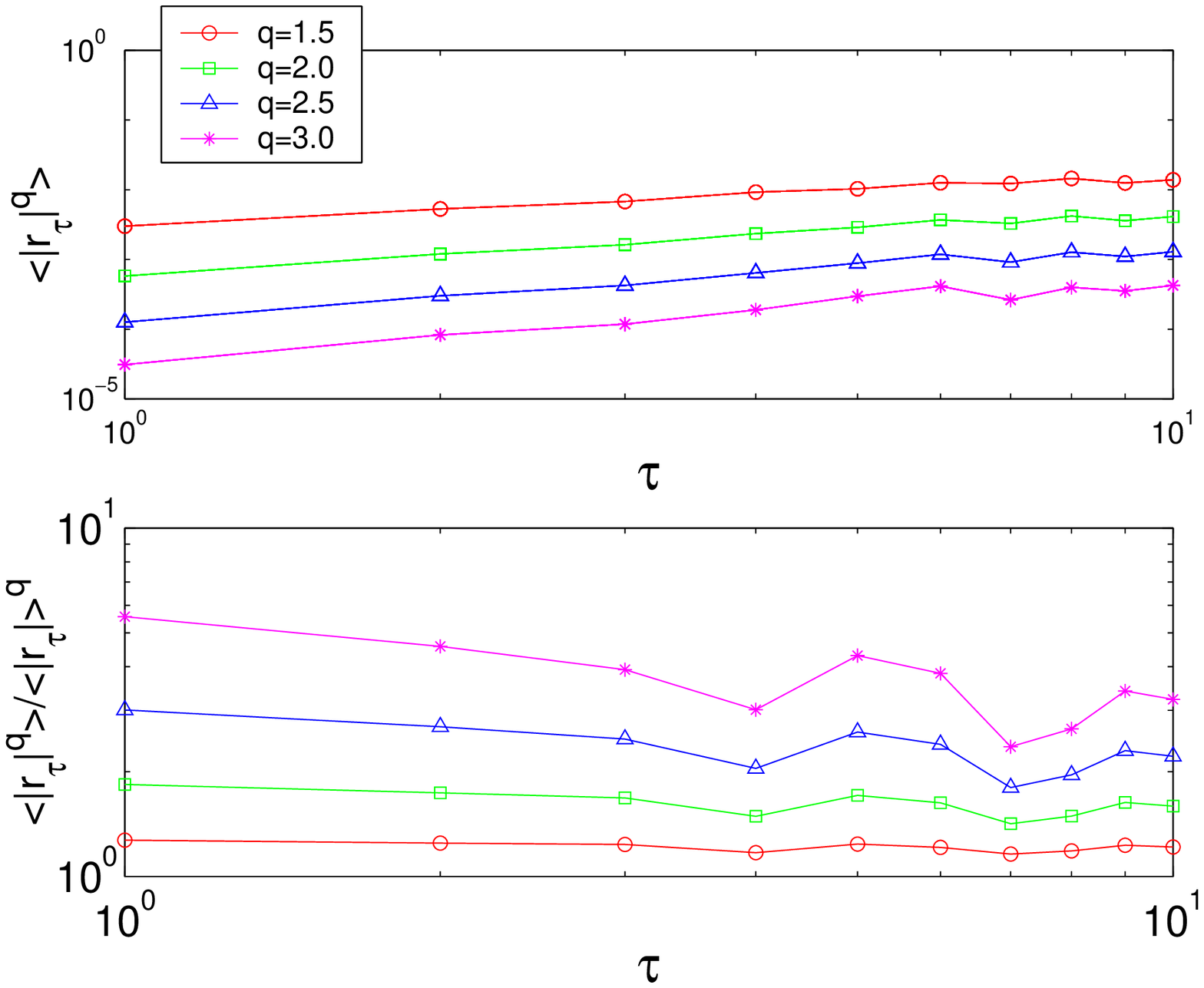} &
\epsfxsize=7.1cm\epsffile{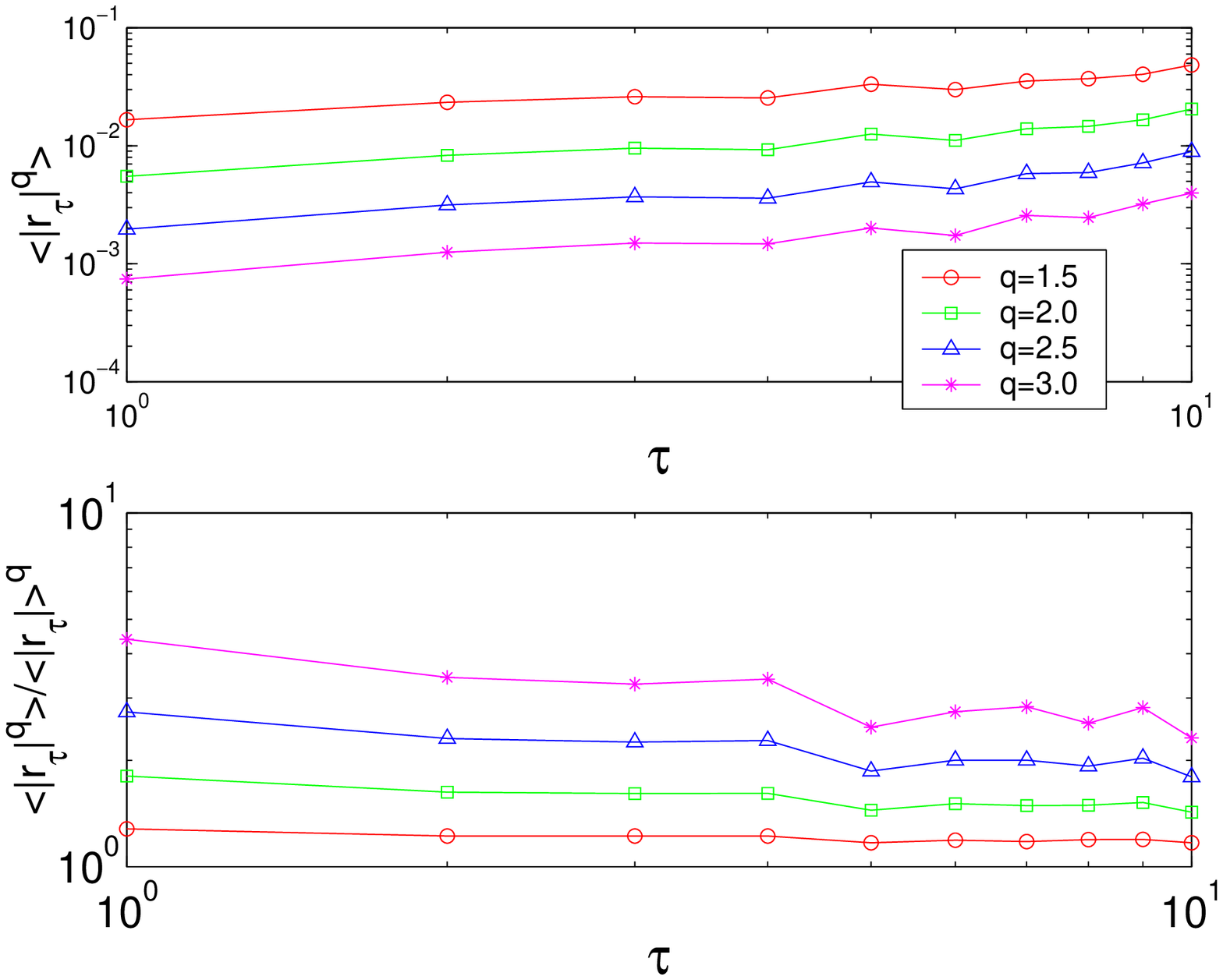} 
\end{tabular}
\caption{Comparison of higher momenta of real (left) and  model data (right).  
Top: The higher momenta of the return process 
are scaling processes with respect to aggregation time $\tau$. 
Bottom: 
$\langle |r_{\tau}|^q \rangle / \langle |r_{\tau}| \rangle ^q  $
is not a constant, indicating a multifractal underlying return process. 
 }
\label{multifrac}
\end{figure}

Finally we computed the higher momenta $\langle |r_{\tau}|^q \rangle$
of the return processes for various $q$ values, 
Fig. \ref{multifrac} (top). 
We observe that $\langle |r_{\tau}|^q \rangle$ scales with 
respect to aggregation time $\tau$. The corresponding log-log 
slopes do not depend on $q$ linearly, which is indicated by a 
slightly declining 
$\langle |r_{\tau}|^q \rangle / \langle |r_{\tau}| \rangle ^q  $, 
Fig. \ref{multifrac} (bottom). 
This suggests that the underlying process is a multifractal. 
When comparing to realistic stock data of about the same data 
length (left),  again,  nice agreement is found. 

The presented model has a rich dynamical structure which will be subject 
for a more detailed investigation in future work. Here we just mention that 
formally the stationary solutions are  given by
$p    =0$, 
$\xi_1=\pm \sqrt{{\cal T}_1} $, and 
$\xi_2=\pm \frac{1}{\tau_1^2} 
( - {\cal T}_0) /  \sqrt{ {\cal T}_1 }$. 
The determinant of the Jacobian matrix is
${\rm Det} (M)= - 2c \,  \xi_1^3  / \tau_2^2 $   
which for the stationary solutions 
reduces to $- 2c \,  {\cal T}_1^{3/2} /  \tau_2^2 $.
For the same set of parameters as before  the attractor is  plotted 
in Fig. \ref{resA}d for several time steps. 
Note that the attractor is on the ``tic-scale'' and that the price 
process used for the above analysis takes only every $1000^{\rm th }$
point (on average) from that trajectory.

\section*{Discussion and Conclusion} 
We have   presented a  non-equilibrium price evolution 
model, which captures  regulatory 
price movements through the agents' desire to use mispricings 
to  make profits,  and which takes into account  
risk-aversion of those agents. 
We contrasted  this model to the 
standard believes of financial economics  where  
prices emerge from equilibrium, by agents maximizing their 
utility functions. In our approach we do not 
use  the standard mathematical tools like Wiener processes 
or  game theory. 
The apparent differences of our dynamical system with the 
standard view is the explicit use of a non-equilibrium concept, 
which is realized by the introduction of thermostats which model 
the behavior of  agents. Our model does not contain any 
sources of randomness, the erratic behavior originates 
from the non-linear  nature of the model. 
Similarities to the standard approach are that we also use the concept 
of  risk-aversion and that the basic price 
equation is formally similar to the standard formulation. 
The presented model - even though  very simple conceptually - 
leads to  realistic looking price dynamics over a wide range of 
parameter settings. 
Volatility clustering, fat-tailed distributions 
of returns, correct looking autocorrelation functions, 
higher momenta and multifractal spectra 
are well reproduced.  
Of course, a purely deterministic model can never be the 
full truth of a complex system like financial markets, but 
it is intriguing that it is possible to explain most stylized facts 
of its resulting time series.  
Whether there are practical sides to the model remains to be 
seen. To mention an apparent consequence,  
in the present framework the pricing of financial derivatives 
would be 
fundamentally different from current practice where 
arbitrage free prices are derived from replicating portfolios, 
which are chosen such that the (Wiener) random sources  
cancel each other. 



\end{document}